\documentclass[a4paper]{jpconf}
\usepackage{graphicx}
\begin{document}
\title{PICO-LON Project for WIMPs search}

\author{Ken-Ichi Fushimi$^{1}$, Kensuke Yasudai$^{1}$, Yuuki Kamedai$^{1}$,
Hiroyasu Ejiri$^{2,3,4}$, Ryuta Hazama$^{5}$, Kayoko Ichihara$^{6}$, 
Kyoshiro Imagawa$^{7}$, Hiroshi Ito$^{7}$, Norihiko Koori$^{1}$, 
Hidehito Nakamura$^{3}$, Shintaro Nakayama$^{1}$, Masaharu Nomachi,$^{6}$, 
Tatsushi Shima$^{2}$, Saori Umehara$^{6}$,  Sei Yoshida$^{8}$}
\address{1. Faculty of Integrated Arts and Sciences, 
The University of Tokushima, Tokushima city Tokushima 770-8502, JAPAN}
\address{2. Research Center for Nuclear Physics, Osaka University, 
Ibaraki city Osaka, 567-0043 JAPAN}
\address{3. National Institute of Radiological Sciences, Chiba city Chiba 
263-8555, JAPAN}
\address{4. Nuclear Science, Czech Technical University, Prague, 
Czech Republic}
\address{5. Faculty of Engineering, Hiroshima University, 
Higashi Hiroshima city Hiroshima 739-8527, JAPAN}
\address{6. Department of Physics, Osaka University, Toyonaka city Osaka, 
560-0043 JAPAN}
\address{7. Horiba Ltd., Kyoto city Kyoto,601-8510, JAPAN}
\ead{kfushimi@ias.tokushima-u.ac.jp}
\address{8. Research Center for Neutrino Science, Tohoku University, 
Sendai city Miyagi 980-8578, JAPAN}

\begin{abstract}
Highly segmented inorganic crystal has been shown to have good performance 
for dark matter search.
The energy resolution of ultra thin and large area NaI(Tl) scintillator 
has been developed.
The estimated sensitivity for spin-dependent excitation of $^{127}$I 
was discussed.
The recent status of low background measurement at Oto Cosmo Observatory 
is reported
\end{abstract}

\section{Introduction}
Dark matter search is one of the most 
important subjects in nuclear- and astro-particle physics.
The particle candidates for cosmic dark matter is a key subject 
not only for astrophysics but also for particle physics since the particle 
candidates for dark matter is proposed by various models 
the beyond standard model.

The components of the universe have been studied by many cosmological
observations..
Since the most of the matter in the universe should be the cold dark matter, 
searching for WIMPs (Weakly Interacting Massive Particles) is 
quite important.
The dark matter in the galaxy has become quite ensuring.
One of the most promising candidate for WIMPs is the SUSY neutralino, which 
interacts with the matter via only weak interaction.

The processes for WIMPs-nucleus interaction are divided into three types; 
spin-independent elastic scattering (SI), spin-dependent elastic
scattering (SD) and spin-dependent inelastic excitation (EX).

In the case of the type-(SI) the scattering amplitude is summed coherently 
over all the nucleons.
Thus the scattering cross section is enhanced by the factor 
$A^{2}$,
where $A$ is the mass number of the target nucleus.
In the case of the type-(SD),  only one nucleon which carries the nuclear spin
contributes to the cross section.
Thus the cross section is a few orders of magnitude smaller than that of 
the SI case. 
The cross section depends on the nuclear spin-matrix element 
$\lambda^{2}J(J+1)$
and has a large ambiguity because 
this matrix element has a large model dependence for the heavy nuclei 
\cite{Ellis}.
In the case of type-(EX), the target nucleus will be excited to the 
low lying excited state, which is followed by the gamma ray emission.
This process is arisen only if the WIMPs particle has enough kinetic energy.
In this case, the matrix element is evaluated from the measured rate of the 
gamma de-excitation\cite{Ellis,Ejiri}. 
Thus the model dependence of the cross section is much smaller than 
the cross section of the SD case.

The segmentation of the detector is shown to be the best way to 
enhance the position sensitivity\cite{FushimiJPSJ74}.
Coincidence measurements of nuclear recoils and $\gamma$ rays for the 
inelastic excitation of $^{127}$I enhanced the sensitivity using 
the highly segmented NaI(Tl) detector.
Recently, ionized atomic electrons and hard X rays following 
WIMPs-nuclear interactions have been shown to be useful for 
the exclusive measurement of
nuclear recoils from the elastic scatterings of WIMPs off nuclei
\cite{ejiri-X,ejiri-X2}.
On the other hand, the background events have their own characteristics of 
timing and spatial profiles.
Because the event rate due to the background is reduced by segmentation, 
a probability of the accidental coincidence of 
individual background events is vastly reduced.

It has been shown that piling up many thin scintillators enhances 
the sensitivity for WIMPs search\cite{FushimiJPSJ74}.
The highly sensitive detector system PICO-LON
(Planar Inorganic Crystals Observatory for LOw-background Neutr(al)ino)
has been developed.
It consists of many thin NaI(Tl) crystals whose thickness is 0.05cm.
Recently, PICO-LON-I which consists of three crystals of 
thin (0.05cm in thickness) and wide area (6.6cm$\times$6.6cm) 
NaI(Tl) crystals has been developed.
PICO-LON-II which consists of sixteen crystals of thin NaI(Tl) has been 
also developed.
In this report, we describe the excellent performance of a single 
plate of thin NaI(Tl) which is the foundation of PICO-LON system.
\section{Performance test and low background measurement}
The energy resolution of the thin NaI(Tl) scintillator was measured for 
low energy gamma rays.
The gamma rays from the sources, $^{241}$Am, $^{133}$Ba and $^{57}$Co
were irradiated without any collimator.
The typical energy resolution (FWHM) was obtained as small as 18\% at 60keV
\cite{FushimiJPSJ75}.

The energy dependence of the energy resolution was measured.
The energy resolution changes inversely proportional to the square root 
of the energy.
The obtained energy dependence agrees with the relationship.
From the relationship, the corresponding energy for the single photoelectron 
was calculated as 0.35keV.

The low background measurement was  started in September 2007 at 
Oto Cosmo Observatory.
The observatory is located in the south of Nara prefecture and 150km 
east from Tokushima.
There are three laboratories at the center of the unused railway tunnel
whose length is 5km, and the thickness of the overburden rock is 450m.
The flux of cosmic rays was reduced by four orders of magnitude compared with 
the flux at the surface observatory.
The fast neutron was reduced by two orders of magnitude.

The concentration of radioactive radon, which is most serious origin of 
the background against dark matter search, is kept as low as a few Bq/m$^{3}$
out of the air-tight container.
The PICO-LON system was installed in the air-tight container in which pure
nitrogen gas was used. 
The concentration of radon in the air-tight container is kept as low as 
150mBq/m$^{3}$.
The huge NaI(Tl) scintillators whose dimension of 
10.6cm$\times$10.6cm$\times$106cm were used as the active shield.
The PICO-LON detector was installed between the huge NaI(Tl) detectors.
The air-tight container was covered with 10cm thick pure copper shield and 
15cm thick old lead shield to reduce the gamma ray background from the
surrounding rock.
\section{Prospect} 
The installation and the tuning of the PICO-LON detector will be done in the 
winter of 2007. The full measurement will be started in the beginning of 2008.
The larger NaI(Tl) plates whose dimension of 15cm$\times$15cm$\times$0.1cm 
for each plate have been developed and installed in the spring of 2008.
The result of high sensitivity measurement will be reported in the summer 
of 2008.

\begin{figure}
\begin{center}
\includegraphics[width=14pc]{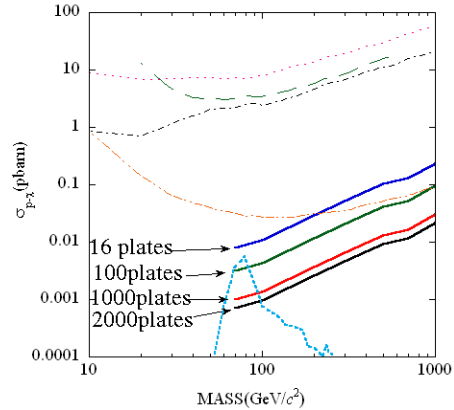}
\end{center}
\caption{\label{fg:limit}Expected sensitivity for spin-dependent excitation.
The cross section was normalized to the proton-WIMPs cross section.
The dotted line, dashed line, dash-dotted line and 
dash-double-dotted lines are the 
experimental obtained by CRESST, DAMA LXe, ELEGANT-V and NAIAD groups.}
\end{figure}

The PICO-LON system has the ability of extension.
The modular plates are piled up and the large volume detector 
system is constructed.
The sensitivity for EX-type of WIMPs interaction depends on the statistics 
of the energy region of interest (57.6keV).
The expected sensitivity by further extension of the PICO-LON system is
shown in figure \ref{fg:limit}.
The expected sensitivities by PICO-LON project are calculated by assuming the 
main background origin is the U-chain contamination which is concentrated in 
the NaI(Tl) crystal.
The concentration of U-chain nuclei are assumed as small as 10$\mu$Bq/kg
for $^{214}$Pb and $^{214}$Bi, 100$\mu$Bq/kg for $^{210}$Pb
\cite{FushimiJPSJ74}.

\section{Acknowledgment}
The authors thank Prof.T.Kishimoto for continuous encouragement of the 
project.
The present work was supported by ``Toray Science and Technology Grant''
and the University of Tokushima.

\end{document}